\documentstyle[prl,aps,floats,psfig]{revtex}

\begin{document}
\twocolumn[
\hsize\textwidth\columnwidth\hsize\csname@twocolumnfalse\endcsname

\draft
\title{Neutron scattering and superconducting order parameter in 
YBa$_2$Cu$_3$O$_7$}

\author{I.~I.~Mazin}
\address{Geophysical Laboratory, Carnegie Institution,
5251 Broad Branch Road, NW, Washington, DC 20015}

\author{Victor M.~Yakovenko}
\address{Department of Physics and Center for Superconductivity,
University of Maryland, College Park, MD 20742-4111}

\date{{\bf E-print cond-mat/9502025}, Phys.~Rev.~Lett.\ {\bf 75}, 4134
(1995); Erratum {\bf 76}, 1984 (1996). }

\maketitle

\begin{abstract}
   We discuss the origin of the neutron scattering peak at 41 meV
observed in YBa$_2$Cu$_3$O$_7$ below $T_c$. The peak may occur due to
spin-flip electron excitations across the superconducting gap which
are enhanced by the antiferromagnetic interaction between Cu spins. In
this picture, the experiment is most naturally explained if the
superconducting order parameter has  $s$-wave symmetry and 
opposite signs in the bonding and antibonding electron bands formed
within a Cu$_2$O$_4$ bilayer.

\end{abstract}

]

   Neutron scattering is an important tool in the study of high-$T_c$
superconductors. A sharp magnetic neutron scattering peak was observed
in the superconducting state of YBa$_2$Cu$_3$O$_x$ at a 2D wave vector
${\bf Q}=(\pi /a,\pi /b)$ ($a$ and $b$ are the lattice spacings of the
CuO$_2$ plane) and energy $\omega $ equal to 41 meV at $x\approx
7$\cite {RM91}. This effect was confirmed by several groups for $x=7$
\cite {Mook93,Fong95} and $6.5<x<7$ \cite{RM94,Tranquada92}. The peak
has the following remarkable features: (a) It appears only below the
superconducting transition temperature $T_c$ \cite{experiment}; (b) It
is localized in both energy and wave vector; (c) It has sinusoidal
dependence on $q_z$, the wave vector perpendicular to the CuO$_2$
planes, which implies perfect antiferromagnetic correlation of the two
planes in a Cu$_2$O$_4$ bilayer.

\begin{figure}
\centerline{\psfig{file=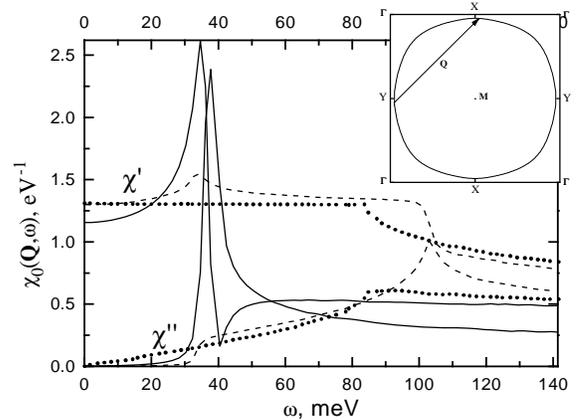,width=0.95\linewidth}}
\caption{The imaginary $\chi^{''(-)}_0({\bf Q},\omega)$  and real
$\chi^{'(-)}_0({\bf Q},\omega)$ parts of the electron spin
susceptibility in the superconducting $s^{\pm}$ (solid and dashed
lines) and normal (dots) states at $T=0$.  Dashed lines and dots
correspond to the Fermi energy $\mu=-370$ meV, solid lines to
$\mu=-440$ meV.  Inset: The Fermi surface for $\mu=-440$ meV.  }
\label{ImRe}
\end{figure}

   A number of theories, suggested to explain this effect, presume
that the peak occurs due to spin-flip electron excitations across the
superconducting gap
\cite{RM91,Fong95,BSS,CTH,BS,MS,Maki,SPL,FKT,LS,ORM}.  It was
emphasized in Ref.\ \cite{Fong95} that, since magnetic scattering is
odd with respect to the time reversal, the BCS coherence factor in the
neutron scattering amplitude vanishes unless $\Delta _{{\bf k}}$ has
opposite signs for the electron wave vectors ${\bf k}$ and ${\bf
k}+{\bf q}$ connected by the 2D neutron wave vector transfer ${\bf q}$
(see the inset to Fig.\ \ref{ImRe}):
\begin{equation}
\Delta _{{\bf k}}\Delta _{{\bf k}+{\bf q}}<0.  \label{DD0}
\end{equation}
Through this condition, neutron scattering can probe the symmetry of
the superconducting state.  For ${\bf q}={\bf Q}$, condition
(\ref{DD0}) is not satisfied for a simple $s$-wave state, but is
satisfied for the $d_{x^2-y^2}$ state. So, it was suggested in Refs.\
\cite{Fong95,CTH,BS,MS,Maki,SPL,FKT,LS,ORM} that the peak in question
is a manifestation of the $d_{x^2-y^2}$ pairing in YBa$_2$Cu$_3$O$_7$.

   However, since most of these theories, as well as Ref.\ \cite{DZ},
dealt with only one CuO$_2$ layer, they were unable to consider
important feature (c). Because YBa$_2$Cu$_3$O$_7$ consists of
Cu$_2$O$_4$ bilayers, there should be two electron bands formed by
bonding and antibonding states. An important issue is the relative
sign of the superconducting order parameter in these two bands. Along
with a regular $d$-wave state which has the same sign in the two
bands, another state, which we will call $s^{\pm }$, was discussed in
the framework of weakly \cite{BSS,LMA,LLM} or strongly correlated
electrons\cite{ILMA,UL,KL}. In this state, the order parameter has
$s$-wave symmetry and opposite signs in the bonding and antibonding
bands. We show below that this state (unlike the $d$-wave state)
provides the best explanation of the neutron scattering experiments,
particularly, of feature (c).

   The neutron scattering cross-section is proportional to the
imaginary part of the electron spin susceptibility $\chi({\bf
q},q_z,\omega)$. In a bilayer system, it is given by the following
expression \cite{Tranquada92,UL}:
\begin{equation}
\chi({\bf q},q_z,\omega)=\chi^{(+)}\cos^2(q_zd/2) + \chi^{(-)}\sin^2(q_zd/2),
\label{Chiz}
\end{equation}
where $d$ is the distance between the CuO$_2$ planes in the bilayer, and 
\begin{eqnarray}
\chi^{(+)}({\bf q},\omega)&=&\chi^{(aa)}({\bf q},\omega) +
\chi^{(bb)}({\bf q},\omega),  \label{chi+} \\
\chi^{(-)}({\bf q},\omega)&=&\chi^{(ab)}({\bf q},\omega) + 
\chi^{(ba)}({\bf q},\omega).  \label{chi-}
\end{eqnarray}
Here, $b$ and $a$ refer to the bonding and antibonding electron bands
of the bilayer. The susceptibilities $\chi^{(ij)}$ account for the
transitions between the respective bands.

   In experiments \cite{RM91,Mook93,Fong95,RM94,Tranquada92}, only the
second term in Eq.\ (\ref{Chiz}) was observed (feature (c)). According
to Eq.\ (\ref {chi+}), the coefficient $\chi ^{(+)}$, which appears in
the first term of Eq.\ (\ref{Chiz}), involves transitions within the
same band. Following the coherence factor arguments, we conclude that
condition (\ref{DD0}) is not satisfied within the same band, that is,
the gap has $s$-symmetry. On the other hand, the coefficient $\chi
^{(-)}$ of the second term of Eq.\ (\ref {Chiz}), which involves
transitions between the different bands (\ref{chi-}), is not
suppressed. Thus, the order parameters of the different bands,
$\Delta^{(a)}$ and $\Delta^{(b)}$, have the opposite signs. This means
that the state is the $s^{\pm }$ state described above. In this
picture, the neutron scattering peak is due to the excitation of
electrons, say, from the bonding band below the superconducting gap to
the antibonding band above the superconducting gap, and the gaps have
opposite signs.

   In order to illustrate the above discussion, we performed explicit
calculations for a well-known realistic tight binding model of
YBa$_2$Cu$_3$O$_7$, which has the following electron dispersion law:
$\xi _{{\bf k}}=-2t(\cos (k_xa)+\cos (k_ya))-4t^{\prime }\cos
(k_xa)\cos (k_ya)-\mu $, where $t=250$ meV, and $t^{\prime }/t=-0.45$
\cite{AJLM}. We have chosen the Fermi energy $\mu =-440$ meV, so that
the van Hove singularity lies at an energy $\xi_{\rm vH}=10$ meV below
the Fermi level \cite{AJLM,Gofron}. The corresponding Fermi surface is
shown in the inset to Fig.\ \ref{ImRe}. To simplify the calculations,
we set the hopping amplitude between the layers, $t_{\perp }$, equal
to zero. This assumption does not qualitatively change our
conclusions.

   In BCS theory, at zero temperature, the susceptibilities are given
by the following formula \cite{Schrieffer}:
\begin{eqnarray}
&&\chi _0^{(ij)}({\bf q},\omega )=\frac 12\sum_{{\bf k}}
\left( 1-\frac{\xi _{{\bf k}+{\bf q}}\xi _{{\bf k}}+
\Delta _{{\bf k}+{\bf q}}^{(i)}\Delta _{{\bf k}}^{(j)}}
{E_{{\bf k}+{\bf q}}E_{{\bf k}}}\right) \times  \nonumber \\
&&\left( \frac 1{\omega +E_{{\bf k}+{\bf q}}+E_{{\bf k}}+i\Gamma }-
\frac 1{\omega -E_{{\bf k}+{\bf q}}-E_{{\bf k}}+i\Gamma }\right),  
\label{chi0}
\end{eqnarray}
where $E_{{\bf k}}=\sqrt{\xi _{{\bf k}}^2+\Delta _{{\bf k}}^2}$ is the
quasiparticle dispersion law in the superconducting state, $\Gamma $
is the damping constant, and the indices $i$ and $j$ label the bonding
and antibonding electron bands. Because of the simplifying assumption
$t_{\perp }=0$, the dispersion laws of the bonding and antibonding
bands are identical and do not need to be distinguished in Eq.\
(\ref{chi0}). Only the signs of the order parameters $\Delta ^{(i)}$
may depend on the index $i$. Note that only when (\ref{DD0}) is
satisfied, does the coherence factor (the first line of Eq.\
(\ref{chi0})) not vanish at small energies $\xi \ll \Delta $.

   We have calculated the spin susceptibility $\chi _0^{(-)}$ directly
from Eqs.\ (\ref{chi0}) and (\ref{chi-}) using $\Gamma =1$ meV and the
$s^{\pm }$ gap computed in Ref.\ \cite{LMA} for $T_c=90$ K. The gap
attains its maximal value $\Delta _0=17.5$ meV near the points X and Y
in the Brillouin zone (see the inset to Fig.\ \ref{ImRe}). Imaginary
$\chi_0^{\prime\prime(-)}({\bf Q},\omega )$ and real
$\chi_0^{\prime(-)}({\bf Q},\omega )$ parts of the susceptibility are
shown in Fig.\ \ref{ImRe} for the superconducting and normal states
for two values of the Fermi energy $\mu$.  We observe that in the
superconducting state, for the energies $\omega $ lower than the
absorption threshold $E_{{\rm th}}\approx 35$ meV (``spin gap''),
$\chi _0^{\prime \prime }(\omega )$ is zero. Furthermore, for the
realistic $\mu =-440$ meV {\it both} real and imaginary parts (solid
curves) have sharp peaks, at 35 meV and 38 meV, respectively. In order
to clarify the origin of these features, we repeated the calculations
(dashed curves) with an unrealistic choice of the Fermi energy $\mu
=-370$ meV, which moves the van Hove singularity much deeper below the
Fermi level: $\xi _{{\rm vH}}=80$ meV. In this case, the peak in $\chi
_0^{\prime }(\omega )$ stays at the same energy $E_{\rm th}$, and
$\chi _0^{\prime\prime }(\omega )$ develops a step at the same energy.
On the other hand, the peak in $\chi _0^{\prime \prime }(\omega )$
shifts to much higher energy of about 100 meV, which is close to
$\Delta_0+\xi_{\rm vH}$. $\chi _0^{\prime }(\omega )$ develops a
negative step at the same energy.  Similar results were obtained in
Ref.\ \cite{LS} for the $d_{x^2-y^2}$ state.

   To gain a qualitative understanding of this behavior, we set
$\Gamma\rightarrow 0$ and set the coherence factor in Eq.\
(\ref{chi0}) to 1. In this approximation, $\chi _0^{\prime\prime
}({\bf q},\omega )$ is proportional to the joint density of states
$A({\bf q},\omega)=\sum_{{\bf k}}\delta (\omega -E_{{\bf k}+{\bf
q}}-E_{{\bf k}})$.  The two-particle energy $E_2({\bf k},{\bf
q})=E_{{\bf k}+{\bf q}}+E_{{\bf k}} $, considered as a function of the
2D wave vector {\bf k} for a fixed {\bf q}, has a minimum and several
saddle points. The minimum defines the threshold energy $E_{{\rm
th}}$.  In a 2D case, the joint density of states has a step at the
threshold and logarithmic divergences at the saddle point
energies. Correspondingly, the real part, $\chi _0^{\prime }(\omega
)$, which is related with $\chi _0^{\prime \prime }(\omega )$ by the
Kramers-Kronig relations, has a logarithmic singularity at the
threshold. Exactly that behavior is observed in Fig.\ \ref{ImRe}.  The
minimum of $E_2({\bf k},{\bf Q})$ is achieved at a vector {\bf k} such
that both {\bf k} and {\bf k}+{\bf Q} belong to the Fermi surface (see
the inset to Fig.\ \ref {ImRe}), where both $E_{{\bf k}}$ and $E_{{\bf
k}+{\bf Q}}$ attain their minimal values approximately equal to
$\Delta_0$. Thus, $E_{{\rm th}}\approx2\Delta_0=38$ meV.

   The logarithmic peak in $\chi _0^{\prime \prime }(\omega )$ occurs
because of transitions between the occupied states located near X and
Y points and empty quasiparticle states above the superconducting
gap. The points X and Y are the saddle points of the {\it
normal-state} dispersion law $\xi _{{\bf k}}$.  They produce the van
Hove singularity in the {\it single-particle} density of states in the
{\it normal} state at the energy $\xi _{{\rm vH}}$. The logarithmic
divergence in the {\it joint} density of states in the {\it
superconducting} state is located, then, at the energy
$E^{*}\approx\Delta _0+\sqrt{\Delta _0^2+\xi _{{\rm vH}}^2}$. The
value of $\xi _{{\rm vH}}$ depends on the Fermi energy $\mu $. For the
two choices of $\mu$ in Fig.\ \ref{ImRe}, the values of $\xi _{{\rm
vH}}$ are equal, respectively, to 10 and 80 meV, which gives values of
$E^{*}\approx 38$ meV and $E^{*}\approx 100$ meV, respectively, in
agreement with the positions of the peaks in $\chi _0^{\prime \prime
}(\omega )$ in Fig.\ \ref{ImRe}. Note that both LDA
calculations\cite{AJLM} and photoemission experiments \cite{Gofron}
place the van Hove singularity very close to the Fermi level, which
corresponds to the first choice.

   Comparison of $\chi _0^{\prime \prime }({\bf Q},\omega )$ in the
normal state at zero temperature (dotted curve in Fig.\ \ref{ImRe})
and in the superconducting state (dashed line) shows that the
logarithmic peak due to the van Hove singularity is absent in the
normal state. Instead, we see a cusp at an energy close to $\xi_{{\rm
vH}}=80$ meV, which marks the threshold where transitions from the
saddle points to the Fermi level become allowed. The reason that,
instead of a cusp in the normal state, a divergence develops in the
superconducting state is that in the latter case, roughly speaking,
the transitions take place between the van Hove singularity and the
coherence peak in the quasiparticle density of states.  Since there is
no threshold of absorption in the normal state, $\chi _0^{\prime
}(\omega )$ has no divergence at a finite $\omega $ in this state
(dotted line in Fig.\ \ref{ImRe}).

   Summarizing, we conclude that the peaks in Fig.\ \ref{ImRe} appear
to manifest two different physical mechanisms. The peak in
$\chi^{\prime}_0({\bf Q},\omega )$ is due to the transitions between
the states just below the gap to the states right above it at the
energy $E_{\rm th}\approx2\Delta_0$ which depends only on the
superconducting gap. On the other hand, the peak in
$\chi^{\prime\prime }_0({\bf Q},\omega)$ is due to the transitions
from an occupied saddle point to the empty states just above the
superconducting gap. The energy of this peak depends on the van Hove
energy of the normal state $\xi_{\rm vH}$.  For a realistic Fermi
energy $\mu =440$ meV, $\xi_{\rm vH}=10$ meV is small compared to
$\Delta_0=17.5$ meV, therefore the peaks in $\chi^{\prime}_0$ and
$\chi^{\prime\prime }_0$ approach each other, which is shown below to
have a major effect on neutron scattering. None of these peaks appear
in the normal state.

\begin{figure}
\centerline{\psfig{file=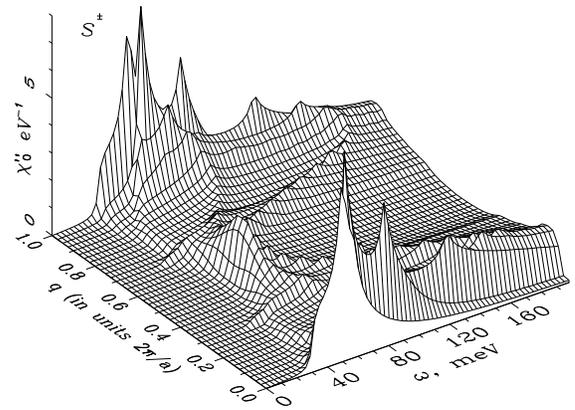,width=0.9\linewidth}}
\caption{Bare susceptibility $\chi^{\prime\prime(-)}_0(q_x,q_y,\omega)$,
plotted as a function of $q=q_x=q_y$ and $\omega$ for the $s^\pm$ order
parameter with $\mu=-440$ meV.}
\label{chi0"}
\end{figure}

   Curves in Fig.\ \ref{ImRe} are reminiscent of the experimental
curves \cite{RM91,RM94}, where $\chi^{\prime \prime }({\bf Q},\omega
)$ was found to exhibit a spin gap and a peak at different energies in
underdoped samples with x$<$7.  It is tempting \cite{LS} to identify
the logarithmic peak in $\chi_0^{\prime\prime }({\bf Q},\omega )$ with
the peak observed in the experiment.  However, the plot of
$\chi_0^{\prime\prime }({\bf q},\omega )$ reveals a structure quite
different from that found experimentally. While there is a local
maximum at ${\bf q}={\bf Q}$ and $\omega\approx 2\Delta _0$ in Fig.\
\ref{chi0"}, the corresponding peaks, although weaker, exist also at
other values of {\bf q}, in contradiction with feature (b).  Even
worse, when {\bf q}$\rightarrow 0$, $\chi _0^{\prime\prime}({\bf
q},\omega )$ increases to much higher values than at ${\bf q}={\bf
Q}$. We conclude that $\chi _0^{\prime \prime }({\bf q},\omega )$ does
not provide a satisfactory explanation of the experimentally observed
neutron scattering.

   Thus far, we have neglected the interaction between quasiparticles
in the superconducting state. If we take into account a weak
antiferromagnetic interaction between Cu spins in the plane, $J({\bf
q})$, and between the planes, $J_{\perp}$, the bare spin
susceptibilities $\chi_0^{(\pm)}$, discussed above, should be
renormalized in an RPA manner \cite {Maki,SPL,FKT,UL,Si}:
\begin{eqnarray}
\chi^{(\pm)}({\bf q},\omega)&=&\chi^{(\pm)}_0({\bf q},\omega)/ 
[1+J^{(\pm)}({\bf q})\chi^{(\pm)}_0({\bf q},\omega)/2],  \label{chi} \\
J^{(\pm)}({\bf q})&=&J({\bf q}) \pm J_{\perp}.  \label{J}
\end{eqnarray}
Since the {\it real} part, $\chi^{\prime(-)}_0({\bf q},\omega)$,
diverges as $\omega\rightarrow E_{{\rm th}}$, as discussed above, the
renormalized susceptibility $\chi^{(-)}({\bf q},\omega)$ (\ref{chi})
has a pole at $\omega $ close to $E_{{\rm th}}$ with singularities in
both real and imaginary parts of $\chi^{(-)}$. Physically, the pole in
$\chi^{(-)}({\bf q},\omega)$ describes a triplet electron-hole
collective mode (an ``antiparamagnon'' or a ``triplet exciton'') with
the energy slightly below $E_{{\rm th}}$. In this picture, the peak in
the neutron scattering rate occurs due to inelastic excitation of this
collective mode by neutrons.

\begin{figure}
\centerline{\psfig{file=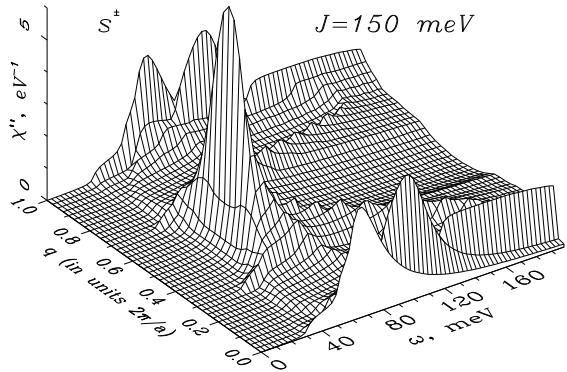,width=0.9\linewidth}}
\caption{Renormalized susceptibility 
$\chi^{\prime\prime(-)}(q_x,q_y,\omega)$, plotted as a function of
$q=q_x=q_y$ and $\omega$ for the same $s^\pm$ state as in Fig.\
\protect\ref{chi0"}.}  
\label{s+-} 
\end{figure}

   Since the singularity in $\chi _0^{\prime (-)}({\bf q},\omega )$
exists, generally speaking, for any {\bf q}, the exciton state (and,
thus, a peak in neutron scattering) should exist for any {\bf q}, in
contradiction with feature (b). However, if we take into account a
finite lifetime of quasiparticles $\Gamma $ in Eq.\ (\ref{chi0}), the
excitons also acquire a finite lifetime. For the antiferromagnetic
interaction between the nearest neighbors, $\chi ^{(-)}({\bf q},\omega
)$ acquires a peak at ${\bf q}={\bf Q}$, where $J({\bf q})=J_{\Vert
}[\cos (q_xa)+\cos (q_yb)]$ reaches its maximal negative value. Thus,
the position of the neutron peak in {\bf q}-space is set by $J({\bf
q})$ and in $\omega $ by $\chi _0^{\prime (-)}({\bf Q},\omega )$. This
statement is illustrated in Fig.\ \ref{s+-}, where we show $\chi
^{\prime \prime }({\bf q},\omega )$ calculated according to Eq.\
(\ref{chi}) with $J^{(-)}=150$ meV and $\Gamma=1$ meV for the $s^{\pm
}$ state. In agreement with experiment, a single peak in $\chi
^{\prime\prime(-)}({\bf q},\omega)$, localized both in {\bf q} and
$\omega$, is observed, which is now more than twice higher than that
at $q \rightarrow 0$. The magnitude of the peak depends on the chosen
value of $J$ and $\Gamma$, but the qualitatively the picture remains
the same for a reasonable range of these parameters.  Similar results
were obtained in Ref.\ \cite{FKT} for the $d_{x^2-y^2}$ state. It is
worth noting that the peak in $\chi^{''(-)}$ is due to the peak in the
real part of $\chi_0^{(-)}$. When the van Hove singularity is close to
the Fermi level, the peak in $\chi _0^{\prime (-)}({\bf Q},\omega ) $
gets stronger and enhances the peak in the renormalized susceptibility
$\chi ^{(-)}({\bf q},\omega )$.

   The crucial difference between the $s^{\pm}$ and $d_{x^2-y^2}$
states is due to the following: In the $s^{\pm}$ state, the divergence
occurs only in $\chi_0^{\prime(-)}$, but not in $\chi_0^{\prime(+)}$
where it is suppressed by the coherence factor (\ref{chi0}). This
suppression, unlike the case of logarithmic divergence in
$\chi_0^{\prime\prime(+)}$, is exact, because the divergence in the
real part of $\chi_0^{(+)}$ would occur precisely at the threshold
energy where the coherence factor vanishes. As a consequence, among
the renormalized susceptibilities, only $\chi^{(-)}$ diverges, but not
$\chi^{(+)}$. Taking in account Eq.\ (\ref{Chiz}), this naturally
explains feature (c) of the experiment.

   In the case of the $d$-wave, the coherence factor allows
divergences in both $\chi _0^{\prime (-)}$ and $\chi _0^{\prime
(+)}$, which, in turn, produce divergences in both $\chi ^{(-)}$ and
$\chi^{(+)}$. The only way to reconcile this with the experiment is to
assume that $J^{(+)}$ is small: $|J^{(+)}|\ll |J^{(-)}|$, or even has
a wrong (positive) sign. According to Eq.\ (\ref{J}), this would
require that the antiferromagnetic interaction between the layers be
stronger than within a layer: $J_{\perp }\gtrsim 2J_{\Vert}$, which is
unrealistic \cite{LLZ}.

   A number of theoretical papers \cite{BSS,CTH,BS,MS,Maki} consider
the special case when $t^{\prime }=0$ in the dispersion law. This
assumption results in nesting at the Fermi energy or at another energy
(``dynamic nesting''), which produces a peak in $\chi _0^{\prime
\prime }({\bf Q},\omega )$. However, this case is not relevant for
YBa$_2$Cu$_3$O$_7$ where {\bf Q} is not a nesting vector.

   In conclusion, we considered a scenario where the peak at 41 meV
observed in neutron scattering experiments
\cite{RM91,Mook93,Fong95,RM94} occurs due to spin-flip interband
electron excitations across the superconducting gap which are enhanced
by the antiferromagnetic interaction between Cu spins. We found that
the experiment can be explained most naturally if the superconducting
order parameter is of the $s^{\pm }$ type, that is, the order
parameter which has the $s$-wave symmetry and the opposite signs in
the bonding and antibonding bands. This state easily explains the
observed dependence of the scattering intensity on the momentum
perpendicular to the CuO$_2$ planes. On the other hand, the
$d_{x^2-y^2}$ case can be reconciled with the observed dependence only
if the antiferromagnetic interaction between the CuO$_2$ planes is
stronger than the interaction within the plane, which seems to be
unrealistic.

   We thank D.~Reznik and B.~Keimer for making their results available
to us prior to publication and for useful discussions. The work of
V.M.Y. was supported in part by NSF under Grant DMR-9417451 and by the
A.~P.~Sloan Foundation.


\begin{references}
\bibitem{experiment}  In Ref.\ \cite{Mook93}, the peak was found to persist
above $T_c$, albeit with greatly reduced intensity.
\end{references}
\end{document}